\documentclass[conference]{IEEEtran}
\usepackage{amsmath}
\usepackage{graphicx}
\usepackage{caption2}
\usepackage{amsthm}

\makeatletter
\renewenvironment{proof}[1][\proofname]{%
  \par\pushQED{\qed}\normalfont%
  \topsep6\p@\@plus6\p@\relax
  \trivlist\item[\hskip\labelsep\bfseries#1\@addpunct{.}]%
  \ignorespaces
}{%
  \popQED\endtrivlist\@endpefalse
}
\makeatother
\begin{document}
\title{A Novel Proof for the DoF Region of the MIMO Broadcast Channel with No CSIT}
\author{\IEEEauthorblockN{Borzoo Rassouli, Chenxi Hao and Bruno Clerckx}
\IEEEauthorblockA{Communication and Signal Processing Group, Department of Electrical and Electronic Engineering\\Imperial College London, United Kingdom\\
Email: \{b.rassouli12; chenxi.hao10; b.clerckx\}@imperial.ac.uk}
}
%\thanks{This work was partially supported by the Seventh Framework Programme for Research of the European Commission under grant number HARP-318489.}}
\maketitle
\begin{abstract}
%\boldmath
In this paper, a new proof for the degrees of freedom (DoF) region of the $K$-user multiple-input multiple-output (MIMO) broadcast channel (BC) with no channel state information at the transmitter (CSIT) and perfect channel state information at the receivers (CSIR)  is provided. Based on this proof, the capacity region of a certain class of MIMO BC with channel distribution information at the transmitter (CDIT) and perfect CSIR is derived. Finally, an outer bound for the DoF region of the MIMO interference channel (IC) with no CSIT is provided.\footnote{This work was partially supported by the Seventh Framework Programme for Research of the European Commission under grant number HARP-318489.}
\end{abstract}

\section{Introduction}
Spatial multiplexing is a key feature of MIMO communication networks \cite{Bruno}. The DoF region, which is the capacity region normalized by the logarithm of SNR in high SNR regimes, is a metric that captures the spatial multiplexing property. The DoF region of the MIMO BC with no CSIT was first shown in \cite{Huang0}, \cite{Huang1} for the two user case and later in \cite{Vaze} for the general $K$-user BC.

In this paper, we provide a new proof for the results obtained in the mentioned papers based on a simple lemma. The paper is organized as follows. Section \ref{system} introduces the system model and the characterization of the DoF region. Our new proof is provided in section \ref{proof}. Based on this proof, the capacity region of a certain $K$-user MIMO BC with CDIT and an outer bound for the DoF region of the MIMO IC with no CSIT are provided in section \ref{capacity} and section \ref{IC}, respectively. Section \ref{conc} concludes the paper.

Throughout the paper, $(.)^H$ and $R_{\geq 0}$ denote the conjugate transpose and the set of non-negative real numbers, respectively. Also, $f\sim o(\log P)$ is equivalent to $\lim_{P\to \infty}\frac{f}{\log P}=0$.
\section{System Model and Main Results}\label{system}
We consider a MIMO BC, in which a transmitter with $M$ antennas sends independent messages $W_1,\ldots,W_K$ to $K$ users (receivers), where each receiver is equipped with $N_i$ receive antennas ($i=1,2,\ldots,K$). In a flat fading scenario, the discrete-time baseband received signal of user $i$ at channel use $t$ can be written as
\begin{equation}\label{e1}
  \textbf{{Y}}_i(t)=\textbf{{H}}_i^H(t)\textbf{{X}}(t)+\textbf{{Z}}_i(t)\ \ \ i=1,2,\ldots,K
\end{equation}
where $\textbf{{Y}}_i(t)\in C^{N_i\times 1}$ is the received signal at receiver $i$, $\textbf{{X}}(t)\in C^{M\times 1}$ is the transmitted signal satisfying the power constraint $E[\|\textbf{{X}}\|^2]\leq P$, $\textbf{{H}}_i(t)\in C^{M\times N_i}$ is the channel matrix of user $i$ and $\textbf{{Z}}_i(t)\in C^{N_i\times 1}$ is the additive white Gaussian noise at receiver $i$. The elements of $\textbf{{H}}_i(t)$ and $\textbf{{Z}}_i(t)$ are independent identically distributed circularly symmetric complex Gaussian random variables with unit variance. These elements are also assumed i.i.d. across the users. Let $\textbf{H}_i^n=\{\textbf{H}_i(1),\textbf{H}_i(2),\ldots,\textbf{H}_i(n)\}$ be the set of channel matrices of user $i$ up to channel use $n$. We assume no channel state information at the transmitter and perfect channel state information at the receiver (CSIR)  i.e., at channel use $n$, user $i$ has perfect knowledge of $\textbf{H}_i^n$.

The rate tuple $(R_1,R_2,\ldots,R_K)$ is achievable if the probability of error in decoding $W_i$ at user $i (i=1,\ldots,K)$ can be made arbitrarily small with sufficiently large coding length. Analysis of the capacity region $C(P)$, which is the set of all the achievable rate tuples, is not always tractable. Instead, we consider the DoF region, which is a simpler metric independent of the transmit power, and is defined as $\{(d_1,\ldots,d_K)|\exists (R_1,R_2,\ldots,R_K)\in C(P) \mbox{\ such that\ } d_i=\lim_{P\to \infty}\frac{R_i}{\log P} \forall i\}$. At very high SNRs, the effect of additive noise can be neglected and what remains is the interference caused by other users' signals. Therefore, the DoF region could also be interpreted as the region constructed by the number of interference-free private data streams that users receive simultaneously per channel use.

\textbf{Theorem 1.} The DoF region of the $K$-user MIMO BC with no CSIT and perfect CSIR is given by
\begin{equation}\label{e2}
 D=\{(d_1,d_2,\ldots,d_K)\in R_{\geq 0}^K| \sum_{i=1}^K\frac{d_i}{r_i}\leq 1\}
\end{equation}
where $r_i=\min\{M,N_i\}$. The region is achieved by orthogonal transmission schemes, such as time sharing across the users.
\section{Proof of the theorem 1}\label{proof}
Unlike \cite{Huang0} and \cite{Huang1}, the proof is not based on the degradedness of the MIMO BC under no CSIT. Without loss of generality, we assume $N_1\geq N_2\geq\ldots\geq N_K$ and we enhance the channel by giving the message of user $i$ to users $i+1, i+2,\ldots, K$. We also assume that each user not only knows its own channel, but also has perfect knowledge of the other users' channels. In other words, perfect global CSIR is assumed. It is obvious that this assumption does not reduce the outer bound which means that the bound with CSIR is inside the bound with global CSIR; however, the achievability is based on only CSIR not global CSIR. According to the Fano's inequality
\begin{equation}\label{e3}
  nR_i \leq I(W_i;\tilde{\textbf{Y}}_i^n|\Omega^n,W_{i+1},\ldots,W_K)+\epsilon_n\ \ \ \ i=1,2,\ldots,K
\end{equation}
where $W_{K+1}=\emptyset$, $\tilde{\textbf{Y}}_i^n=\{\textbf{Y}_i(1),\textbf{Y}_i(2),\ldots,\textbf{Y}_i(n)\}$ is the extension of the received signal at user $i$ over $n$ channel uses and $\Omega^n=\{\textbf{H}_1^n,\textbf{H}_2^n,\ldots,\textbf{H}_K^n\}$ is the global channel state information up to channel use $n$. We decompose the received observation of user $i$ as $\tilde{\textbf{Y}}_i^n=(\textbf{Y}_i^n,\hat{\textbf{Y}}_i^n)$ where $\textbf{Y}_i^n$ is the set of $r_i$ linearly independent observations and $\hat{\textbf{Y}}_i^n$ can be reconstructed by linear combination of the elements in $\textbf{Y}_i^n$ within noise level. From the chain rule of mutual information,
\begin{align}\label{e4}
  nR_i &\leq I(W_i;\textbf{Y}_i^n|\Omega^n,W_{i+1},\ldots,W_K)\nonumber \\
  &\ \ \ +I(W_i;\hat{\textbf{Y}}_i^n|\Omega^n,W_{i+1},\ldots,W_K,\textbf{Y}_i^n)+\epsilon_n \nonumber \\
  &=I(W_i;\textbf{Y}_i^n|\Omega^n,W_{i+1},\ldots,W_K)\nonumber\\
  &\ \ \ +\underbrace{h(\hat{\textbf{Y}}_i^n|\Omega^n,W_{i+1},\ldots,W_K,\textbf{Y}_i^n)}_{o(\log P)}\nonumber\\&\ \ \ -\underbrace{h(\hat{\textbf{Y}}_i^n|\Omega^n,W_{i},\ldots,W_K,\textbf{Y}_i^n)}_{o(\log P)}+\epsilon_n\ \ i=1,2,\ldots,K.
\end{align}
For simplicity, we ignore $\epsilon_n$ and the terms with $o(\log P)$ and write
\begin{align}
\sum_{i=1}^K\frac{nR_i}{r_i}  &\leq  \sum_{i=1}^K\frac{I(W_i;\textbf{Y}_i^n|\Omega^n,W_{i+1},\ldots,W_K)}{r_i} \nonumber \\
 &\leq \underbrace{\frac{h(\textbf{Y}_K^n|\Omega^n)}{r_K}}_{\leq n\log P} +\sum_{i=1}^{K-1}\left[\frac{h(\textbf{Y}_i^n|\Omega^n,W_{i+1},\ldots,W_K)}{r_i}- \right.\nonumber\\
   &\left.\  \ \frac{h(\textbf{Y}_{i+1}^n|\Omega^n,W_{i+1},\ldots,W_K)}{r_{i+1}}\right] \label{e5}
\end{align}
where we have used the fact that $\frac{h(\textbf{Y}_1^n|\Omega^n,W_{1},\ldots,W_K)}{r_1}\sim o(\log P)$, since with the knowledge of $\Omega^n,W_{1},\ldots,W_K$, the observation $\textbf{Y}_1^n$ can be reconstructed within noise distortion.
Before going further, the following lemma, which is an extension of lemma 1 in \cite{Borzoo}, is needed.

\textbf{Lemma}. Let $\Gamma_N=\{Y_1,Y_2,\ldots,Y_N\}$ be a set of $N(\geq2)$ arbitrary random variables and $\Psi_i^{j}(\Gamma_N)$ be a sliding window of size $j$ over $\Gamma_N$ ($1\leq i,j \leq N$) starting from $Y_i$ i.e.,
\[\Psi_i^{j}(\Gamma_N) = Y_{(i-1)_N+1},Y_{(i)_N+1},\ldots,Y_{(i+j-2)_N+1}\]
where $(.)_N$ defines the modulo $N$ operation. Then,
%\begin{multline}\label{e6}
 % (N-m)h(Y_1,Y_2,\ldots,Y_N|A)\leq \sum_{i=1}^{N}h(\Psi_i^{N-m}(\Gamma_N)|A)\nonumber\\
 %  1\leq m\leq N-1
%\end{multline}
\begin{multline}\label{e6}
  (N-m)h(Y_1,Y_2,\ldots,Y_N|A)\leq \sum_{i=1}^{N}h(\Psi_i^{N-m}(\Gamma_N)|A) \\
  1\leq m\leq N-1
\end{multline}
where $A$ is an arbitrary condition.
\begin{proof}
The lemma can be proved in two ways, either by showing that for every fixed $N(\geq 2)$, (\ref{e6}) holds for all $m$ satisfying $1\leq m\leq N-1$, or by showing that for every fixed $m(\geq 1)$, (\ref{e6}) holds for all $N(\geq m+1)$.  We choose the latter approach and prove it by induction. It is obvious that for every $m(\geq 1)$, (\ref{e6}) holds for $N=m+1$. In other words, $h(Y_1,Y_2,\ldots,Y_N|A)\leq \sum_{i=1}^{N}h(Y_i|A)$. Now, considering that (\ref{e6}) is valid for $N(\geq m+1)$, we show that it also holds for $N+1$. Replacing $N$ with $N+1$, we have
\begin{align}
  &(N+1-m)h(Y_1,\ldots,Y_N,Y_{N+1}|A)\nonumber\\&=\!h(Y_1,\ldots,Y_N,Y_{N+1}|A)+\!(N-m)h(Y_1,Y_2,\ldots,Y_{N-1},\overbrace{Y_N,Y_{N+1}}^{Z}|A)\nonumber\\
  &\leq h(Y_1,\ldots,Y_{N+1}|A)+\sum_{i=1}^{N}h(\Psi_i^{N-m}(\Phi_N)|A)\label{e8}\\
  &= h(Y_1,\ldots,Y_{N+1}|A)+\sum_{i=1}^{m}h(\Psi_i^{N-m}(\Phi_N)|A)\nonumber\\&\ \ \ + \sum_{i=m+1}^{N}h(\Psi_i^{N-m}(\Phi_N)|A)\label{e9}\\
  &= h(Y_1,\ldots,Y_{N+1}|A)+\sum_{i=1}^{m}h(\Psi_i^{N-m}(\Phi_N)|A)\nonumber\\&\ \ \ + \sum_{i=m+1}^{N}h(\Psi_i^{N+1-m}(\Gamma_{N+1})|A)\label{e10}\\
  &= h(Y_{N-m+1},\ldots,Y_{N}|Y_{N+1},Y_1,Y_2,\ldots,Y_{N-m},A)\nonumber\\&\ \ \ +\sum_{i=1}^{m}h(\Psi_i^{N-m}(\Phi_N)|A)+h(Y_{N+1},Y_1,Y_2,\ldots,Y_{N-m}|A)\nonumber\\
  &\ \ \ + \sum_{i=m+1}^{N}h(\Psi_i^{N+1-m}(\Gamma_{N+1})|A)\label{e11}\\
  &= h(Y_{N-m+1},\ldots,Y_{N}|Y_{N+1},Y_1,Y_2,\ldots,Y_{N-m},A)\nonumber\\
  &\ \ \ +\sum_{i=1}^{m}h(\Psi_i^{N-m}(\Phi_N)|A)+\sum_{i=m+1}^{N+1}h(\Psi_i^{N+1-m}(\Gamma_{N+1})|A)\label{e12}\\
  &= h(Y_{N-m+1},\ldots,Y_{N}|Y_{N+1},Y_1,Y_2,\ldots,Y_{N-m},A)\nonumber\\
  &\ \ \ +\sum_{i=1}^{m}h(Y_i,Y_{i+1},\ldots,Y_{N-m+i-1}|A)+\sum_{i=m+1}^{N+1}h(\Psi_i^{N+1-m}(\Gamma_{N+1})|A)\label{e13}\\
  &= \sum_{i=1}^mh(Y_{N-m+i}|Y_{N+1},Y_1,Y_2,\ldots,Y_{N-m+i-1},A)\nonumber\\
  &\ \ \ +\sum_{i=1}^{m}h(Y_i,Y_{i+1},\ldots,Y_{N-m+i-1}|A)+\sum_{i=m+1}^{N+1}h(\Psi_i^{N+1-m}(\Gamma_{N+1})|A)\label{e13.5}\\
  &\leq \sum_{i=1}^mh(Y_{N-m+i}|Y_i,Y_{i+1},\ldots,Y_{N-m+i-1},A)\nonumber
  \end{align}
\begin{align}
&\ \ \ +\sum_{i=1}^{m}h(Y_i,Y_{i+1},\ldots,Y_{N-m+i-1}|A)\nonumber\\&\ \ \ +\sum_{i=m+1}^{N+1}h(\Psi_i^{N+1-m}(\Gamma_{N+1})|A)\label{e13.75}\\
  &= \sum_{i=1}^{m}h(\Psi_i^{N+1-m}(\Gamma_{N+1})|A)+\sum_{i=m+1}^{N+1}h(\Psi_i^{N+1-m}(\Gamma_{N+1})|A) \label{e14}\\
  &= \sum_{i=1}^{N+1}h(\Psi_i^{N+1-m}(\Gamma_{N+1})|A)
\end{align}
where in (\ref{e8}), $\Phi_N=\{Y_1,Y_2,\ldots,Y_{N-1},Z\}$ and we have used the validity of (\ref{e6}) for $N$. In (\ref{e10}), we have used the fact that $\Psi_i^{N+1-m}(\Gamma_{N+1})=\Psi_i^{N-m}(\Phi_N)$ for $i \in [m+1,N]$ . In (\ref{e11}), the chain rule of entropies is used and in (\ref{e13}), the sliding window is written in terms of its elements. Finally, in (\ref{e13.75}), the fact that conditioning reduces the differential entropy is used. Therefore, since $m(\geq 1)$ was chosen arbitrarily and (\ref{e6}) is valid for $N=m+1$ and from its validity for $N(\geq m+1)$ we could show it also holds for $N+1$, we conclude that (\ref{e6}) holds for all values of $m$ and $N$ satisfying $1\leq m\leq N-1$. \qedhere
\end{proof}
It is obvious that lemma 1 in \cite{Borzoo} is a special case of the above lemma for $m=1.$ Each term in the summation of (\ref{e5}) can be written as
\begin{align}
  &\frac{h(\textbf{Y}_i^n|\Omega^n,W_{i+1},\ldots,W_K)}{r_i}- \frac{h(\textbf{Y}_{i+1}^n|\Omega^n,W_{i+1},\ldots,W_K)}{r_{i+1}}=\nonumber\\
  &\frac{r_{i+1}h(\textbf{Y}_i^n|\Omega^n,W_{i+1},\ldots,W_K)\!-\!r_ih(\textbf{Y}_{i+1}^n|\Omega^n,W_{i+1},\ldots,W_K)}{r_ir_{i+1}}\label{e15.75}
\end{align}
\begin{align}
&= \frac{r_{i+1}h(\textbf{Y}_{i,1}^n,\textbf{Y}_{i,2}^n,\ldots,\textbf{Y}_{i,r_i}^n|\Omega^n,W_{i+1},\ldots,W_K)}{r_ir_{i+1}}\nonumber\\
&\ \ \ -\frac{\!r_ih(\textbf{Y}_{i+1}^n|\Omega^n,W_{i+1},\ldots,W_K)}{r_ir_{i+1}}\\
&\leq \frac{\sum_{p=1}^{r_i}h(\Psi_p^{r_{i+1}}(\Gamma_{r_i})|\Omega^n,W_{i+1},\ldots,W_K)}{r_ir_{i+1}}\nonumber\\
&\ \ \ -\frac{r_ih(\textbf{Y}_{i+1}^n|\Omega^n,W_{i+1},\ldots,W_K)}{r_ir_{i+1}} \label{e16}
\end{align}
\begin{align}
  &= \sum_{p=1}^{r_i}\left[\frac{h(\Psi_p^{r_{i+1}}(\Gamma_{r_i})|\Omega^n,W_{i+1},\ldots,W_K)}{r_ir_{i+1}}\right.\nonumber\\&\ \ \left.-\frac{h(\textbf{Y}_{i+1}^n|\Omega^n,W_{i+1},\ldots,W_K)}{r_ir_{i+1}} \right]\label{e17}\\
  &= \sum_{p=1}^{r_i}\left[\frac{h(\textbf{A}_{p,i,n}\textbf{X}^n+\textbf{B}_{p,i,n}|\Omega^n,W_{i+1},\ldots,W_K)}{r_ir_{i+1}}\right.\nonumber\\&\ \ \left.-\frac{h(\textbf{C}_{i,n}\textbf{X}^n+\textbf{D}_{i,n}|\Omega^n,W_{i+1},\ldots,W_K)}{r_ir_{i+1}} \right]\label{e17.5}\\
  &=0 \label{e18}
\end{align}
where in (\ref{e16}), since $r_{i+1}\leq r_i$, the result of the previous lemma is applied in which $\Gamma_{r_i}=\{\textbf{Y}_{i,1}^n,\textbf{Y}_{i,2}^n,\ldots,\textbf{Y}_{i,r_i}^n\}$ is the set of $r_i$ linearly independent elements in $\textbf{Y}_i^n$. In (\ref{e17.5}), we write $\Psi_p^{r_{i+1}}(\Gamma_{r_i})$ and $\textbf{Y}_{i+1}^n$ as large $nr_{i+1}$ dimensional vectors as follows. $\Psi_p^{r_{i+1}}(\Gamma_{r_i})=\textbf{A}_{p,i,n}\textbf{X}^n+\textbf{B}_{p,i,n}$ and $\textbf{Y}_{i+1}^n=\textbf{C}_{i,n}\textbf{X}^n+\textbf{D}_{i,n}$ where $\textbf{A}_{p,i,n}$ and $\textbf{C}_{i,n}$ ($\in C^{nr_{i+1}\times nM}$) capture the channel coefficients over the $n$ channel uses, $\textbf{X}^n$ is the $nM$ dimensional input vector and $\textbf{B}_{p,i,n}$ and $\textbf{D}_{i,n}$ capture the noise vectors over the $n$ channel uses. Since $\textbf{A}_{p,i,n}$ and $\textbf{C}_{i,n}$ are identically distributed channel coefficients and $\textbf{B}_{p,i,n}$ and $\textbf{D}_{i,n}$ are identically distributed noise terms, the arguments of the differential entropies in (\ref{e17.5}) are statistically equivalent (i.e., have the same probability density function). Since the entropies are only a function of the distribution, we conclude that the two entropies in the difference are equal which results in (\ref{e18}). Therefore, (\ref{e5}) is simplified to
\begin{equation}
  \sum_{i=1}^K\frac{nR_i}{r_i}\leq n\log P.
\end{equation}
After dividing both sides by $n\log P$ and taking the limit $n,P \to \infty$, we get
\begin{equation}
  \sum_{i=1}^K\frac{d_i}{r_i}\leq 1.
\end{equation}
The above DoF region is achieved by a simple time sharing across the users where the global CSIR assumption is not necessary.

\textbf{Remark 1}. The DoF region remains unchanged under the assumption of different noise distributions across the users.

In this case, (\ref{e18}) does not hold anymore, since the terms in the differential entropies are no longer statistically equivalent due to different noise distributions. In this case, we further enhance the channel by giving all the noise vectors to all the users. Therefore, (\ref{e4}) is modified as
\begin{align}
  nR_i &\leq I(W_i;\textbf{Y}_i^n,\Lambda^n|\Omega^n,W_{i+1},\ldots,W_K)\nonumber\\&=I(W_i;\textbf{Y}_i^n|\Omega^n,\Lambda^n,W_{i+1},\ldots,W_K)\nonumber\\&\ \ \ +\underbrace{I(W_i;\Lambda^n|\Omega^n,W_{i+1},\ldots,W_K)}_{=0}
\end{align}
where $\Lambda^n$ denotes the set of all the noise vectors across the users (extended over $n$ channel uses). Following the same approach, (\ref{e17.5}) is modified as
\begin{align}\label{e22}
&\sum_{p=1}^{r_i}\left[\frac{h(\textbf{A}_{p,i,n}\textbf{X}^n+\textbf{B}_{p,i,n}|\Omega^n,\Lambda^n,W_{i+1},\ldots,W_K)}{r_ir_{i+1}}\right.\nonumber\\
&\ \ \left.-\frac{h(\textbf{C}_{i,n}\textbf{X}^n+\textbf{D}_{i,n}|\Omega^n,\Lambda^n,W_{i+1},\ldots,W_K)}{r_ir_{i+1}} \right].
\end{align}
The matrices $\textbf{A}_{p,i,n}$ and $\textbf{C}_{i,n}$, which contain the channel coefficients, have the same distribution, however the vectors $\textbf{B}_{p,i,n}$ and $\textbf{D}_{i,n}$, which contain the noise terms, are no longer statistically equivalent. Hence, by taking the expectation over all the noise realizations, (\ref{e22}) becomes
\begin{align}\label{e23}
&\sum_{p=1}^{r_i}\left(E_{\Lambda^n}\left[\frac{h(\textbf{A}_{p,i,n}\textbf{X}^n+\textbf{B}_{p,i,n}|\Omega^n,\Lambda^n=\lambda^n,W_{i+1},\ldots,W_K)}{r_ir_{i+1}}\right.\right.\nonumber\\&\left.\left.-\frac{h(\textbf{C}_{i,n}\textbf{X}^n+\textbf{D}_{i,n}|\Omega^n,\Lambda^n=\lambda^n,W_{i+1},\ldots,W_K)}{r_ir_{i+1}}\right]\right)
\end{align}
where $\lambda^n$ is a realization of $\Lambda^n$. By applying the realization to the arguments of the differential entropies, (\ref{e23}) becomes
\begin{align}\label{e24}
&\sum_{p=1}^{r_i}\left(E_{\Lambda^n}\left[\frac{h(\textbf{A}_{p,i,n}\textbf{X}^n+B_{p,i,n}|\Omega^n,\Lambda^n=\lambda^n,W_{i+1},\ldots,W_K)}{r_ir_{i+1}}\right.\right.\nonumber\\&\left.\left.-\frac{h(\textbf{C}_{i,n}\textbf{X}^n+D_{i,n}|\Omega^n,\Lambda^n=\lambda_n,W_{i+1},\ldots,W_K)}{r_ir_{i+1}}\right]\right)
\end{align}
\begin{align}
  &= \sum_{p=1}^{r_i}\left(E_{\Lambda^n}\left[\frac{h(\textbf{A}_{p,i,n}\textbf{X}^n|\Omega^n,\Lambda^n=\lambda^n,W_{i+1},\ldots,W_K)}{r_ir_{i+1}}\right.\right.\nonumber\\&\ \ \left.\left.-\frac{h(\textbf{C}_{i,n}\textbf{X}^n|\Omega^n,\Lambda^n=\lambda_n,W_{i+1},\ldots,W_K)}{r_ir_{i+1}}\right]\right)\label{e25} \\
  &=0\label{e26}
\end{align}
where $B_{p,i,n}$ and $D_{i,n}$ are the realizations for $\textbf{B}_{p,i,n}$ and $\textbf{D}_{i,n}$, respectively. In (\ref{e25}), we have used the fact that constant addition does not change the differential entropies, and in (\ref{e26}), statistical equivalence between the arguments of the entropies is used. Therefore, the region in the theorem 1 is still an outer bound for the DoF region under the assumption of different noise distributions and since it is achievable, it is still the optimal DoF region in this case. The only difference is in the achievability i.e., since the noise can be non-Gaussian, the Gaussian distribution may no longer be optimal for the input and the optimal input distribution depends on the distribution of the noise in such a way that conditioned on the realization of the channel, the received signal becomes Gaussian.

\textbf{Remark 2}. It is obvious that the assumptions of 1) Gaussian channel distribution and 2) independent channels across the users, were not used in the proof. It means that the proof can also be applied to other correlated channel distributions as long as the channel distributions are identical across the users.
\section{Capacity region analysis}\label{capacity}
In this section we consider the simplest assumptions in the beginning of section \ref{system} i.e., i.i.d. Gaussian channels and noise vectors. We also assume $M\geq N_1\geq N_2\geq\ldots\geq N_K$ which results in $r_i = N_i (i=1,\ldots,K)$. Since the SNR is not necessarily infinite (in contrast to the DoF analysis), all the $o(\log P)$ terms should be replaced with their exact values. The first one is the term in (\ref{e4}) which is zero here, since $M\geq N_1\geq N_2\geq\ldots\geq N_K$ and therefore, $\tilde{\textbf{Y}}_i^n=\textbf{Y}_i^n$. From the Fano's inequality,
\begin{align}
\sum_{i=1}^K\frac{nR_i}{r_i}  &\leq  \sum_{i=1}^K\frac{I(W_i;\textbf{Y}_i^n|\Omega^n,W_{i+1},\ldots,W_K)}{r_i} \nonumber \\
 &\leq \frac{h(\textbf{Y}_K^n|\Omega^n)}{r_K}- \underbrace{\frac{h(\textbf{Y}_1^n|\Omega^n,W_{1},\ldots,W_K)}{r_1}}_{n\log (2\pi e)}+ \nonumber
\end{align}
\begin{equation}\label{e29}
  \underbrace{\sum_{i=1}^{K-1}\!\left[\!\frac{h(\textbf{Y}_i^n|\Omega^n,W_{i+1},\ldots,W_K)}{r_i}-\!\!\right.
   \left. \frac{h(\textbf{Y}_{i+1}^n|\Omega^n,W_{i+1},\ldots,W_K)}{r_{i+1}}\!\!\right]}_{\leq 0}\!\!.
\end{equation}
From the above results, we get an outer bound for the achievable rate region as
\begin{equation}\label{e27.5}
  \sum_{i=1}^K\frac{R_i}{r_i}\leq \frac{h(\textbf{Y}_K^n|\Omega^n)}{nr_K}.
\end{equation}
Therefore, an outer bound for the ergodic capacity region is
\begin{equation}\label{e27.75}
  \sum_{i=1}^K\frac{R_i}{r_i}\leq \frac{\max_{\mathbf{\Sigma}_{X}:\mbox{tr}(\mathbf{\Sigma}_{X})\leq P}E\left[\log \det(\textbf{I}_{r_K}+\textbf{H}_K^H\mathbf{\Sigma}_X\textbf{H}_K)\right]}{r_K}
\end{equation}
and since the channels have i.i.d. Gaussian elements, the optimal input covariance matrix is $\frac{P}{M}\textbf{I}_M$ \cite{Telatar}. Hence,
\begin{align}\label{e28}
  C^o(P)=\{&(R_1,R_2,\ldots,R_K)\in R_{\geq 0}^K| \nonumber \\&R_i\leq E\left[\log \det(\textbf{I}_{r_i}+\frac{P}{M}\textbf{H}_i^H\textbf{H}_i)\right]\ \forall i\nonumber\\&\sum_{i=1}^K\frac{R_i}{r_i}\leq \frac{E\left[\log \det(\textbf{I}_{r_K}+\frac{P}{M}\textbf{H}_K^H\textbf{H}_K)\right]}{r_K}\}
\end{align}
It is obvious that the outer bound is more affected by the capacity of the point-to-point link from the transmitter to the user with the lowest number of receive antennas.

\textbf{Definition.} We define a class of channels (a set of matrices) $\Theta(p,q,m)$ where each channel (matrix) in this class has its elements drawn from the distribution $p$ in such a way that the optimal input covariance matrix for achieving the capacity of the point-to-point link from the transmitter to the virtual user defined by this channel is diagonal with equal entries. The details for this condition are given in \cite[Exercise 8.6]{Tse}. We also assume that for each channel in this class, all the singular values have the distribution $q$. In other words,
\begin{align}
  &\Theta(p,q,m)=\left\{H\in C^{m\times n}\ \forall n|\mbox{ Elements of }H\sim p,\nonumber\right.\\&\arg\max_{\mathbf{\Sigma}_{X}:\mbox{tr}(\mathbf{\Sigma}_{X})\leq P}E\left[\log \det(\textbf{I}_{n}+H^H\mathbf{\Sigma}_XH)\right]=\frac{P}{m}\textbf{I}_m , \nonumber\\&\mbox{and }\lambda_i(H^HH)\sim q, \forall i=1,\ldots,\mbox{rank}(H)\}.
\end{align}
\textbf{Theorem 2.} In a $K$-user Gaussian MIMO BC with $M\geq N_1\geq N_2\geq\ldots\geq N_K$ and all the channels from the class of $\Theta(p,q,M)$, the capacity region with CDIT is given by
\begin{align}\label{e30}
  C(P)=\{&(R_1,R_2,\ldots,R_K)\in R_{\geq 0}^K|\nonumber\\&\sum_{i=1}^K\frac{R_i}{r_i}\leq E_q\left[\log (1+\frac{P}{M}\lambda)\right]\}
\end{align}
where $E_q\left[\log (1+\frac{P}{M}\lambda)\right]=\int \log (1+\frac{P}{M}x)q(x)dx.$
\begin{proof}
According to (\ref{e27.75}) and the properties of $\Theta(p,q,M)$, we have
\begin{equation}
\sum_{i=1}^K\frac{R_i}{r_i}\leq \frac{\sum_{i=1}^{r_K}E\left[\log (1+\frac{P}{M}\lambda_i(\textbf{H}_K^H\textbf{H}_K))\right]}{r_K}.
\end{equation}
If the singular values of $\textbf{H}_K$ have the same distribution, we can write
\begin{equation}
\sum_{i=1}^K\frac{R_i}{r_i}\leq E\left[\log (1+\frac{P}{M}\lambda_1(\textbf{H}_K^H\textbf{H}_K))\right].
\end{equation}
Also, if the singular values have the same distribution across the users, the outer bound is easily achieved by orthogonal transmission strategies, and therefore it is the optimal capacity region.
\end{proof}
A special case of theorem 2 was shown for the two user Gaussian MIMO BC in \cite{Huang1}, in which all the eigenvalues of $\textbf{H}_k^H\textbf{H}_k (k=1,2)$ are unity.
\section{mimo interference channel with no csit}\label{IC}
Consider a $K$-user MIMO IC with $K$ transmitters and $K$ receivers equipped with $M_i$ and $N_i$ antennas, respectively ($i=1,2,\ldots,K$). The input-output relationship at channel use $t$ is given by
\begin{equation}
  \textbf{{Y}}_i(t)=\sum_{j=1}^K\textbf{{H}}_{i,j}^H(t)\textbf{{X}}_j(t)+\textbf{{Z}}_i(t)\ \ \ i=1,2,\ldots,K
\end{equation}
where $\textbf{{Y}}_i(t)$ is the received signal at receiver $i$, $\textbf{{H}}_{i,j}$ is the channel matrix from the transmitter $j$ to the receiver $i$, $\textbf{{X}}_j(t)$ is the transmitted vector by the transmitter $j$ satisfying $E[\|\textbf{X}_j\|^2]\leq P$ and $\textbf{{Z}}_i(t)$ is the noise vector at the receiver $i$. We assume that the channels are drawn from the same distribution, while the noise vectors could have different distributions. We also assume perfect CSIR (each receiver knows all the incoming channels to it from all the transmitters) and no CSIT.
\subsection{2-user MIMO IC}
For the two user case, theorems 2 and 3 in \cite{Huang1} are combined into theorem 5 in \cite{Vaze}. Here, we provide an alternative proof for it. We assume $N_1\leq N_2$ and $r_i=\min(M_2,N_i)$. By giving the message of user 1 to user 2, we have
\begin{align}
  \frac{nR_1}{r_1}+\frac{nR_2}{r_2} &\leq \frac{I(W_1;\tilde{\textbf{Y}}_1^n|\Omega^n)}{r_1}+\frac{I(W_2;\tilde{\textbf{Y}}_2^n|\Omega^n,W_{1})}{r_2}\label{e40}  \\
  &= \frac{h(\tilde{\textbf{Y}}_1^n|\Omega^n)}{r_1}-\overbrace{\frac{h(\tilde{\textbf{Y}}_2^n|\Omega^n,W_{1},W_2)}{r_2}}^{o(\log P)} \nonumber \\
  &\ \ \ +\frac{h(\tilde{\textbf{Y}}_2^n|\Omega^n,W_{1})}{r_2}-\frac{h(\tilde{\textbf{Y}}_1^n|\Omega^n,W_{1})}{r_1}\\
  &\leq \frac{n\min(N_1,M_1+M_2)}{r_1}\log P \nonumber \\
  &\ \ \ +\underbrace{\frac{r_1h(\textbf{Y}_2^n|\Omega^n,W_{1})-r_2h(\textbf{Y}_1^n|\Omega^n,W_{1})}{r_1r_2}}_{\leq 0}\label{e42}\\
  &\leq \frac{n\min(N_1,M_1+M_2)}{r_1}\log P
\end{align}
where in (\ref{e40}), $\tilde{\textbf{Y}}_1^n$ and $\tilde{\textbf{Y}}_2^n$ are the same as those in (\ref{e3}) and we have neglected all the terms with $o(\log P)$ henceforth. In (\ref{e42}), $h(\tilde{\textbf{Y}}_1^n|\Omega^n)$ is maximized when $\tilde{\textbf{Y}}_1^n$ is Gaussian received from a transmitter with $M_1+M_2$ antennas. Also, in the term $[\frac{h(\tilde{\textbf{Y}}_2^n|\Omega^n,W_{1})}{r_2}-\frac{h(\tilde{\textbf{Y}}_1^n|\Omega^n,W_{1})}{r_1}]$, since the entropies are conditioned on $W_1$, $\textbf{X}_1(1),\textbf{X}_1(2),\ldots,\textbf{X}_1(n)$ are known. Therefore, the extensions of $\textbf{H}_{11}^H(t)\textbf{X}_1(t)$ and $\textbf{H}_{21}^H(t)\textbf{X}_1(t)$ over $n$ channel uses can be removed from $\tilde{\textbf{Y}}_1^n$ and $\tilde{\textbf{Y}}_2^n$, respectively. What remains is a broadcast channel with a transmitter having $M_2$ transmit antennas. With a difference of $o(\log P)$, we can replace $\tilde{\textbf{Y}}_1^n$ and $\tilde{\textbf{Y}}_2^n$ with their linearly independent elements $\textbf{Y}_1^n$ and $\textbf{Y}_2^n$, respectively as in (\ref{e4}). Since $r_1\leq r_2$, following the same approach as in the formulae (\ref{e15.75}) to (\ref{e18}), we get the non-positive term in (\ref{e42}). Therefore, the outer bound is
\begin{align}
  D^o=\{&(d_1,d_2)\in R_{\geq 0}^2|\ d_i\leq \min(M_i,N_i)\ i=1,2\mbox{ and }\nonumber\\
  &\frac{d_1}{r_1}+\frac{d_2}{r_2}\leq\frac{\min(N_1,M_1+M_2)}{r_1}\}.
\end{align}
\subsection{$K$-user MIMO IC}
It is obvious that an outer bound for the DoF region of the MIMO IC can be obtained if the transmitters cooperate to make a broadcast channel with $M_T=\sum_{i=1}^KM_i$ antennas at the base station. Following the same proof in this paper for the broadcast channel, we get
\begin{align}
  D^o=\{&(d_1,d_2,\ldots,d_K)\in R_{\geq 0}^K|\nonumber\\&d_i\leq \min(M_i,N_i)\ \forall i\mbox{ and } \sum_{i=1}^K\frac{d_i}{\min(M_T,N_i)}\leq 1\}.
\end{align}
According to theorem 9 in \cite{Vaze}, the above outer bound is tight provided that either $N_i\leq M_i\ \forall i$ or $N_i=N \geq M=M_i \ \forall i$ where in the former time sharing across the users and in the latter receive zero-forcing and time sharing are the achievable schemes, respectively.
\section{Conclusion}\label{conc}
In this paper, a novel proof for the DoF region of the $K$-user MIMO BC with no CSIT was provided. Motivated by the proof, the capacity region of a specific class of the $K$-user Gaussian MIMO BC with CDIT is derived. Also, an outer bound for the DoF region of the MIMO IC with no CSIT is provided.
\bibliography{REFERENCE1}
\bibliographystyle{IEEEtran}
\end{document}